\documentstyle[twocolumn,eqsecnum,aps]{revtex}
\begin{document}
\draft
\title{Calculating the $Q\bar{Q}$ Potential in 2+1 Dimensional
Light-Front QCD}
\author{M. Burkardt and B. Klindworth}
\address{Department of Physics\\
New Mexico State University\\
Las Cruces, NM 88003-0001\\U.S.A.}
\maketitle
\begin{abstract}
Pure glue QCD is formulated on a 2+1 dimensional transverse lattice, 
using discrete light-front quantization. The transverse component
of the gauge fields is taken to be compact, but in a linearized 
approximation with an effective potential. 
The (rest-frame) $Q\bar{Q}$ potential is evaluated numerically
using Lanzcos matrix diagonalization. We first discuss the strong 
coupling limit analytically and then present numerical results 
beyond the strong coupling limit. The physical origin of confinement
on a transverse lattice depends on the orientation of the external
charges: For longitudinally separated charges, confinement arises
from the instantaneous Coulomb interaction and for transversely
separated quarks a string of link-fields forms. In the general case
one obtains a superposition of both effects.
Despite the asymetry in the microscopic mechanism, already a very 
simple ansatz for the effective link-field potential provides an 
almost rotationally invariant $Q\bar{Q}$ potential.
The momentum carried by the glue depends strongly on the orientation
of the external charges, which might have observable consequences.
\end{abstract}
\narrowtext
\section{Introduction}
Light-cone coordinates \cite{dirac} are the natural coordinates for describing
high-energy scattering \cite{xdj,all:lftd,dgr:elfe,mb:adv}. The immense wealth of data on nucleon and
nucleus structure functions thus strongly motivates to understand
QCD on the light-cone. The transverse lattice formulation
of QCD\cite{bardeen,paul,pauldoubl,paul:zako} is a particularly promising approach towards
this goal. Among the most appealing features of this approach
to light-front QCD is
that confinement emerges naturally in the limit of large lattice
spacing\cite{bardeen,mb:elfe}. 
\begin{figure}
\unitlength.45cm
\begin{picture}(15,9.3)(0.,5.5)
\put(1.5,8.5){\line(0,1){1.7}}
\put(1.,11.1){\makebox(0,0){(discrete)}}
\put(1.,12.0){\makebox(0,0){$\perp$ space}}
\put(1.5,12.6){\vector(0,1){1.6}}
\put(2.,8.){\line(3,-1){1.2}}
\put(4.5,6.9){\makebox(0,0){long. space}}
\put(4.5,6.2){\makebox(0,0){(continuous)}}
\put(7.1,6.3){\vector(3,-1){1.8}}
\put(10.4,5.8){\line(3,1){1.8}}
\put(14.5,6.2){\makebox(0,0){(continuous)}}
\put(14.5,6.9){\makebox(0,0){time}}
\put(15.8,7.6){\vector(3,1){1.8}}
\includegraphics{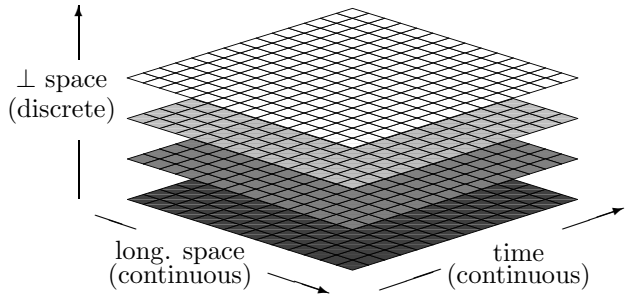}
\end{picture}
\caption{Space time view of a transverse lattice}
\label{fig1}
\end{figure}

In the transverse lattice formulation,
the transverse space directions are discretized, while the longitudinal
[i.e. the $x^\pm \equiv (x^0\pm x^3)/\sqrt{2}$] directions are kept
continuous (Fig.\ref{fig1}). 
While this seems to be a natural procedure when
quantizing on the light-front, the procedure is obviously not
manifestly rotationally invariant and one might ask oneself
whether rotationally invariance is recovered in the continuum limit\cite{mbrot}.
This issue becomes even more relevant, when one considers
"practical" calculations (in contrast to the infinitely
complicated continuum limit), i.e. calculations where the
lattice spacing is not necessarily infinitesimally small
because of numerical limitations or where one makes
approximations on top of the discretization.

In this paper, we will consider one specific observable, namely
the rest-frame potential energy of an (infinitely heavy)
$Q\bar{Q}$ pair coupled to the gluon system.
As has been shown in Ref.\cite{mb:conf}, this observable can be
extracted from a light-front Hamiltonian by considering a
$Q\bar{Q}$ pair which moves with uniform velocity and where
the separation between the quark and the antiquark is kept fixed
(the corresponding formalism is briefly summarized in the Appendix).

After setting the general formalism, we will proceed to calculate
the $Q\bar{Q}$ potential away from the strong coupling limit.
The main issues there are to find out whether linear
confinement persists and whether rotationally invariance for
the potential is being restored.

\section{The Hamiltonian}
The light-front Hamiltonian for compact QCD on a transverse
lattice has been introduced in Ref.\cite{bardeen}. For
pure glue QCD in 2+1 dimensions one finds
\begin{eqnarray}
P^-&=&c_g \sum_n\int dx^-\int dy^-
:\mbox{tr}\left[ J_n(x^-)\left| x^--y^- \right| J_n(y^-)\right]
:\nonumber\\
& &+V_{eff}(U),
\label{eq:pminus}
\end{eqnarray}
where
\begin{equation}
J_n=U^\dagger_n \stackrel{\leftrightarrow}{\partial} U_n
-U^\dagger_{n+1}\stackrel{\leftrightarrow}{\partial} U_{n+1}
\end{equation}
and $U_n$ are the link fields, which are quantized matrix fields and
satisfy the usual
commutation relations. Ideally, one would like to work with
$U_n\in SU(N)$, but in practice this is very complicated\cite{paul:zako}
so in practice one may prefer to work with an unconstrained complex
matrix field and instead add an effective constraint term
$V_{eff}(U)$ to the light-front Hamiltonian. In the case of
$N\rightarrow \infty$, in the classical limit, 
$V_{eff}(U)$ can be taken of the form
\begin{equation}
V_{eff}^{cl}(U)=c_2 \sum_n\mbox{tr}:\left[U^\dagger_n U_n\right]:
+ c_4 \sum_n\mbox{tr}:\left[ U^\dagger_nU_nU^\dagger_nU_n\right]:,
\label{eq:constr}
\end{equation}
where $c_2=-2c_4$ and $c_4\rightarrow \infty$, which
provides an effective
potential that is minimized for $U_n\in U(N)$. In the $N\rightarrow
\infty$ limit, the difference between $U(N)$ and $SU(N)$ is irrelevant
and Eq.(\ref{eq:constr})
is thus suitable for enforcing the $SU(N)$ constraint in the classical
limit. One might be tempted to try a similar ansatz for the
LF quantized case. There are several reasons why a different form for the
effective potential is more useful.
First, if one still attempts to work with an effective
potential of the above form then physical states would
necessarily look extremely complicated, which has to do with the fact
that the above ansatz for the effective potential corresponds to 
working close to the continuum limit. Thus even if the ansatz
in Eq.(\ref{eq:constr}) would work in principle, it would most
likely not be very practical. 

However, since a "Mexican hat"
potential corresponds to a situation where one is working with
the false vacuum, it is questionable whether a physical
situation where a particle runs at the bottom of a Mexican hat
can be described at all by a LF Hamiltonian, using degrees of freedom
expanded around the origin.

For these two reasons, it makes more sense not to consider
$U_n$ as the {\it bare} link field, but instead think of it as
some kind of {\it blocked} or {\it smeared} variable. The blocking
has several consequences. First, the $SU(N)$ constraint gets relaxed which
reflects itself in the fact that the effective potential is no longer
just a narrow valley \cite{pirner}. Secondly, using smeared variables,
it might be easier to cover large physical distances with only
few degrees of freedom. 
The price one has to pay for these advantages is that the
effective potential gets more complex and in general more terms are
necessary than shown in Eq.(\ref{eq:constr}). In Ref.\cite{bvds}
an attempt has been made to fit the effective potential to the
glueball spectrum by making an ansatz which includes all operators
up to dimension four.
Since this work is a first study of the rest-frame $Q\bar{Q}$ potential
and, as we will discuss below, the $Q\bar{Q}$ potential turns out
to be rather insensitive to terms of dimension greater than two
in $V_{eff}(U)$, we will instead only consider a much simpler ansatz
in the following and keep only the quadratic term
\begin{equation}
V_{eff}(U) \approx c_2 \mbox{tr}\left(U^\dagger U\right). 
\label{eq:vc2}
\end{equation}
\section{Numerical Procedure and Results}
\label{sec:num}
Once one has made an ansatz for the Hamiltonian, one needs to find
approximate solutions. In this work, we will employ DLCQ \cite{pa:dlcq}
for this purpose, i.e. on each link the gauge field $U_n$ will be
expanded in of a plane wave basis of complex matrix fields.
For simplicity, zero mode degrees of freedom will be omitted in this
procedure, but we impose a color singlet constraint at each site
\begin{equation}
Q_n|phys\rangle \equiv \int dx^- J_n(x^-)|phys\rangle =0,
\label{eq:zero}
\end{equation}
which is known to results after eliminating zero mode degrees of
freedom in the limit of a large longitudinal interval \cite{engel}.
\footnote{Note that even though this has been shown only for
1+1 dimensional gauge theories in Ref.\cite{engel}, the result can be
used here since QCD on a  transverse lattice is formally equivalent to
a large number of coupled 1+1 dimensional field theories!}
In the large $N$ limit, the color singlet constraint is easily
satisfied by using a basis of states that can be written as
traces over color indices. This has been shown in Ref.\cite{bardeen}
and will be used in this paper.

To some extend, one can justify the omission of zero modes in
the sense of Refs.\cite{ho:vac,dgr,mbsg,fr:eps,brazil,mb:parity} 
by formulating
the theory in terms of an effective
light-front Hamiltonian, where high-energy degrees of freedom
have been integrated out. While this procedure certainly takes 
care of some of the zero-mode dynamics, it is not clear whether
this allows one to completely omit zero-modes as dynamical degrees
of freedom. Therefore, one should regard the omission of zero-modes
(as dynamical degrees of freedom) in this paper as an {\it ad hoc}
approximation which is applied to make the problem simpler.
Including dynamical zero-modes is one of the many possible
improvements that one could consider as an extension of this
work \cite{zero}.

In principle, there can be an arbitrary number of gluon quanta on
each of the links of the transverse lattice. The only constraint is
abovementioned
color singlet requirement on each site. In order to simplify the
numerical calculation, we will first 
calculate the $Q\bar{Q}$ potential 
under the (ad hoc)
approximation that each link contains at most one
gluon quantum. In combination with the color singlet
requirement this implies that a $Q\bar{Q}$ pair
separated by N transverse links will be connected
by exactly N gluons --- one on each links in between
\footnote{Without this approximation, this would
be the minimal configuration, i.e. any physical state
subject to the color singlet constraint at each site
will have at least this number of gluons.}.
The reason for doing this approximation is that it allows to illustrate
the confinement mechanism on the LF more clearly. Further below we
will show results that do not make use of this approximation.

The physical meaning of 
the {\it one gluon per link} 
approximation is as follows:In the limit of large gluon masses
(=large lattice spacings) a  $Q\bar{Q}$ pair is connected
by a static (no fluctuations) chain of gluons. As the
gluon mass gets smaller, the string starts fluctuating.
Fluctuations where the string moves longitudinally
with respect to the straight line connecting the  $Q\bar{Q}$
pair correspond to excitations which do not change the
number of link fields. Those are included in the above
approximation. Fluctuations which are excluded in the
above approximation are the ones where the string
gets deformed transversely so strongly that it
winds forth and back in the transverse direction.
If one wants an exact solution to QCD, all fluctuations must be 
included. However, as a first step beyond the (static !) heavy gluon 
limit, it makes sense to include only the "motion" of the gluons first 
but not creation and annihilation of additional gluons.
Note that if one were to work close to the continuum limit then
such an approximation would not make sense. However, since we
consider the $U$'s as blocked variables, which correspond to rather
rigid degrees of freedom, it is not completely unreasonable to
assume that pair creation is suppressed in ground state configurations.

The general formalism for calculating the (rest-frame)
$Q\bar{Q}$ potential within the light-front framework
has been discussed in detail in Ref.\cite{mb:conf} and
the reader in strongly encouraged to consult these
works for details. In the following we will only present the 
explicit integral equations that one obtains for the coupled
$Q\bar{Q}-glue$ system.
If the quark and the anti-quark are on the same site,
with longitudinal separation $x^-$,the calculation is trivial 
since the "integral equation"collapses into one single equation.
\begin{equation}
P^- = G^2 \frac{\pi}{2}|x^-|,
\label{eq:0site}
\end{equation}
where $G^2=g^2C_F/\pi$ and $C_F=(N^2_C-1)/2N_C$.
$g$ carries dimensions of mass. In the naive continuum limit, 
is related to the coupling constant $g_0$ in the
Lagrangian of $QCD_{2+1}$ via $g^2 \propto g_0^2/a$, where $a$ is the
transverse lattice spacing, which is why
the coupling has this unusual dimension. As has been shown in 
Ref.\cite{mb:conf},
$x^-$ is related to the longitudinal separation of the
$Q\bar{Q}$ pair in its rest frame $x_L$ via 
\begin{equation}
x_L=x^-v^+,
\label{eq:xl}
\end{equation}
where $v^\mu$ is the velocity vector of the $Q\bar{Q}$
pair. Similarly, $P^-$ is related to the potential
interaction energy between the quark and the anti-quark
(again in their rest frame) through the relation
\cite{mb:conf}
\begin{equation}
V(x_L,x_\perp) = P^-v^+,
\label{eq:v}
\end{equation}
i.e. for zero transverse separation one obtains
\begin{equation}
V(x_L,0)=G^2\frac{\pi}{2}|x_L|.
\label{eq:lstring}
\end{equation}
As soon as the $Q\bar{Q}$ pair is separated by at least
one transverse site, the result is less trivial.
For example, when the transverse separation is one
lattice spacing, one obtains an integral equation
for the wave function of the (one !) gluon connecting
the $Q\bar{Q}$ pair
\begin{eqnarray}
P^-\psi(k^+) &=& \left( \frac{k^+}{2{v^+}^2}+
\frac{m^2}{2k^+}\right)\psi(k^+)
\label{eq:1site} 
\\
& &\!\!\!\!\!\!\!\!\!\!\!\!\!\!\!\!\!\!\!\!\!\!\!\!\!\!\!\!\!\!\!+\, G^2\!
\int_0^\infty \!\!\!\!dq^+
\frac{\left(q^++k^+\right)\left[\psi(k^+)-
\psi(q^+)e^{i(k^+-q^+)x^-/2 }\right]}
{2\sqrt{k^+q^+}\left(q^+-k^+\right)^2}
\nonumber
\\
& &\!\!\!\!\!\!\!\!\!\!\!\!\!\!\!\!\!\!\!\!\!\!\!\!\!\!\!\!\!\!\!+\, G^2\!
\int_0^\infty \!\!\!\!dq^+
\frac{\left(q^++k^+\right)\left[\psi(k^+)-
\psi(q^+)e^{i(q^+-k^+)x^-/2}\right]}
{2\sqrt{k^+q^+}\left(q^+-k^+\right)^2}                                           ,
\nonumber
\end{eqnarray}
where the two interaction terms arise from the Coulomb coupling 
of the gluon to the quark and anti-quark respectively.
Eqs.(\ref{eq:xl}) and (\ref{eq:v}) hold also here. 
The exponential "form -factors" in the interaction terms arise since
quark and anti-quark are displaced in the longitudinal direction.

In the continuum
(i.e. when one solves this integral equation exactly)
the resulting $V(x_L,x_\perp)$ is independent of the
velocity $v^+$. Note that this is not the case when one
uses DLCQ to solve the integral equation. This point will 
be discussed below.

For a separation of two sites the integral
equation for the two gluons between the charges reads
\begin{eqnarray}
& &\!\!\!\!\!P^-\psi(k^+_1,k^+_2) = \left(\frac{k^+_1+k^+_2}{2{v^+}^2}+
\frac{m^2}{2k^+_1}+\frac{m^2}{2k^+_2}\right)
\psi(k^+_1,k^+_2)
\nonumber\\
&+& G^2\!\!\!\int_0^\infty \!\!\!\!dq^+
\frac{\!\left(q^++k^+_1\right)\!\!\left[\psi(k^+_1,k^+_2)-
\psi(q^+,k^+_2)e^{i(k_1^+-q^+)x^-/2 }\right]}
{2\sqrt{k_1^+q^+}\left(q^+-k^+_1\right)^2}
\nonumber\\
&+& G^2\!\!\!\int_0^\infty \!\!\!\!dq^+
\frac{\!\left(q^++k^+_2\right)\!\!\left[\psi(k^+_1,k^+_2)-
\psi(k^+_1,q^+)e^{i(q^+-k_2^+)x^-/2 }\right]}
{2\sqrt{k_2^+q^+}\left(q^+-k^+_2\right)^2}
\nonumber\\
&+& G^2\!\!\int_0^{k_1^++k_2^+} 
\!\!\!\!dq^+
\frac{\left(q^++k^+_1\right)
\left(k_1^++2k_2^+-q^+\right)}
{4\sqrt{k_1^+k_2^+q^+(k_1^++k_2^+-q^+)}}\nonumber\\
& &\quad \quad \quad \quad \quad\quad \quad \quad \quad \quad
\times\frac{\left[\psi(k^+_1,k^+_2)-\psi(q^+,k^+_2)\right]}
{\left(q^+-k^+_1\right)^2}
\nonumber\\
&+& G^2\frac{\pi}{4\sqrt{k_1k_2}}\psi(k^+_1,k^+_2)
\label{eq:2site}
\end{eqnarray}
The three interaction terms in Eq.(\ref{eq:2site})
arise from the interaction of the first gluon with
the quark, the second gluon with the anti-quark and
the interaction between the two gluons, respectively.

The generalization of these expressions to more than
2 links is straightforward but the resulting expressions
are very lengthy and will be omitted here.

Note that we used, the "Coulomb trick"\cite{simon} in
Eqs.(\ref{eq:1site}) and (\ref{eq:2site}) by adding
and subtracting analytically a term in the interaction. This results
in an interaction that vanishes for constant wave functions ---
which is close to the actual shape --- and thus numerical convergence  
is improved considerably:
\begin{eqnarray}
& &\int_0^1\frac{dy}{\left(x-y\right)^2}
\left\{\frac{\left(x+y\right) \left(2-x-y\right)}{\sqrt{xy(1-x)(1-y)}}
\psi(y)-4\psi(x)\right\}
\nonumber\\
& &=
\int_0^1\frac{dy}{\left(x-y\right)^2}
\left\{\frac{\left(x+y\right) \left(2-x-y\right)}{\sqrt{xy(1-x)(1-y)}}
\left[\psi(y)-\psi(x)\right]\right\}
\nonumber\\
& &\quad-\frac{\pi\psi(x)}{\sqrt{x(1-x)}} + \frac{4\psi(x)}{x(1-x)}
,
\label{eq:coulomb}
\end{eqnarray}
as well as
\begin{eqnarray}
& &\int_0^\infty\frac{dy}{\left(x-y\right)^2}
\left\{\frac{\left(x+y\right)}{\sqrt{xy}}
\psi(y)-2\psi(x)\right\}
\nonumber\\
& &=
\int_0^\infty\frac{dy}{\left(x-y\right)^2}
\left\{\frac{\left(x+y\right)}{\sqrt{xy}}
\left[\psi(y)-\psi(x)\right]\right\}
\nonumber\\
& &\quad + \frac{2\psi(x)}{x}
,
\label{eq:coulomb2}
\end{eqnarray}

Even though these integral equations will be solved below
in their fully relativistic form, it is very instructive
to consider various approximations thereof --- particularly
the limit $m^2 \rightarrow \infty$. Making a non-relativistic
expansion around the minimum of the kinetic terms one finds
for the (rest frame) potential energy for an arbitrary configuration 
with gluon (rest frame) positions at $x_1,...,x_{n_\perp}$
\begin{eqnarray}
V = n_\perp m &+& G^2\frac{\pi}{2}\left[|-\frac{x_L}{2}-x_1|+|x_{n_\perp}-\frac{x_L}{2}|
\right.\nonumber\\
& &\left.\quad\quad\quad\quad +\sum_{i=1}^{n_\perp-1} |x_i-x_{i+1}|
\right].
\end{eqnarray}
Clearly, this expression is minimized for\\
$-\frac{x_L}{2}<x_1<...<x_{n_\perp}<\frac{x_L}{2}$ with minimum value
\begin{equation}
V(x_L,x_\perp) = n_\perp m + G^2\frac{\pi}{2}|x_L|.
\label{eq:vstatic}
\end{equation}
For large $m$ this is the $Q\bar{Q}$ potential in the rest frame of
the pair, which thus exhibits linear confinement\cite{mb:conf}. 
Since the lattice spacing has not yet been fixed so far,
we are free to choose
\begin{equation}
a = \frac{2m}{\pi G^2},
\end{equation}
which renders the longitudinal and the transverse
string tension equal to each other. Nevertheless,
$V$ in Eq.(\ref{eq:vstatic}) is still not rotationally
invariant: for example, for $x_L=x_\perp=x$ one finds
$V(x,x)=2 V(0,x)$. For a rotationally invariant linear
potential the result would have been $V(x,x)=\sqrt{2}V(0,x)$.

Such a result is familiar from the strong coupling limit 
in Euclidean or Hamiltonian lattice QCD, where one obtains
exactly the same square shaped equipotential lines. Of course,
once one no longer restricts oneself to the strong coupling
limit, the quantum mechanical fluctuations tend to restore
rotationally invariance.

This is also what happens here. To see this, let us consider
a $Q\bar{Q}$ pair separated by one transverse lattice
unit. When the (rest frame) positions of the quark and the
anti-quark are both at $x_L=0$ then the gluon in between them
experiences a potential energy equal to
\begin{equation}
V_{aligned}(x_1) = G^2\pi \left|  x_1 \right|  .
\label{eq:align}
\end{equation}
In contrast, when the positions of the quark and the
anti-quark are at $\pm \frac{x_L}{2}\neq 0$
the potential energy which the gluon in between them
experiences is
\begin{eqnarray}
V_{displaced}(x_1)&=&\frac{G^2\pi}{2}\left[
\left|-\frac{x_L}{2}-x_1\right| +\left| \frac{x_L}{2}-x_1
\right| \right]
\nonumber\\[2.ex]
&=& \left\{\begin{array}{ll} G^2\pi |x_1| & \mbox{\,\,for\,\,}
|x_1|>\frac{|x_L|}{2}\\[1.ex]
G^2\pi \frac{|x_L|}{2} & \mbox{\,\,for\,\,} |x_1|<\frac{|x_L|}{2} \end{array}\right.
\label{eq:displ}.
\end{eqnarray}
The violation of rotationally invariance in the large $m$ limit
manifests itself through
the fact that the minimum of the potential (w.r.t. $x_1$) 
for the displaced case is too high compared to the aligned case.
\begin{figure}
\unitlength1.cm
\begin{picture}(15,7.)(1.7,-9.5)
\includegraphics{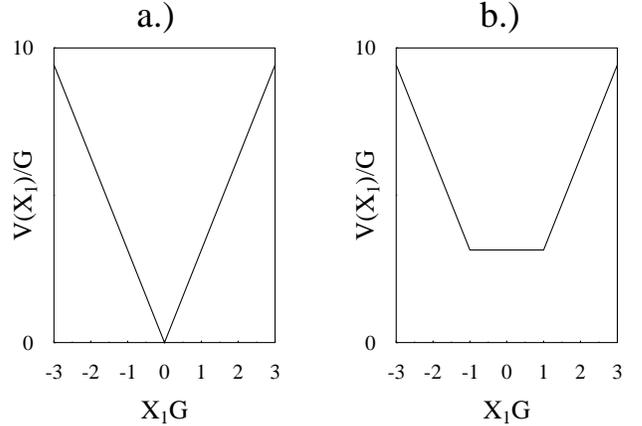}
\end{picture}
\caption{Potential energy of a gluon between an external
$Q\bar{Q}$ pair separated by one link in the (fictitious)
non-relativistic limit. a.) for $Q$ and $\bar{Q}$ at the same
longitudinal position. b.) for $Q$ and $\bar{Q}$ displaced in
the longitudinal direction by $x_L=2/G$.
}
\label{fig:vnr}
\end{figure}

The crucial point is ( see Fig.\ref{fig:vnr})
that the valley of the potential in the displaced case
[Eq.(\ref{eq:displ})] is wider than the valley of the potential in the aligned case 
[Eq.(\ref{eq:align})]. Therefore, corrections due to the quantum mechanical 
zero point energy tend to increase the energy more strongly in the
aligned case than in the displaced case. In a sense, quantum mechanical
corrections work in the right direction to help restore rotationally invariance.

Even though this simple quantum mechanical argument is very useful
in order to understand the physics of the restoration of rotationally
invariance, we will not attempt to make it quantitative since
we will now make a fully relativistic calculation, where we
actually solve light-front integral equations for these systems,
Eqs. (\ref{eq:1site}), (\ref{eq:2site}) and the generalization
to more sites.

As we have seen above, the mass term plays an important role in setting
the scale for the transverse string tension. In fact, given the
string tension $\sigma$, one obtains
$a = \sigma/m$ for large $m^2$.
Thus, if one wants to be close to the continuum limit, one should
try to make $m^2$ as small as possible.
It turns out that below $m^2=0$, the spectrum becomes tachyonic
\cite{mb:bvds},
i.e. $m^2=0$ is the smallest meaningful value. As we will see below,
the string tension in lattice units (i.e. also the lattice spacing in
physical units) remains finite at $m^2=0$. Since $m^2=0$ is thus the
closest we can get to the continuum limit within our approximation,
we will focus only on this value. 

In order to fix the scales, we first determine the transverse
string tension. There are two possibilities to determine this
observable. One is to consider the $Q\bar{Q}$ potential for
large separations of the pair. The other is to consider
a torus geometry, where the energy of glueball states which
"wrap around" the torus is equal to the string tension times
the circumference of the circle\cite{bvds}. Even though the
latter method cannot be used to calculate the $Q\bar{Q}$ potential
or to calculate the string tension in any direction other than
the transverse direction, it turns out to converge faster
than calculations of the $Q\bar{Q}$ potential. Therefore,
"wrap-around glueballs" were used to fix the string tension
in this work.

\begin{figure}
\unitlength1.cm
\begin{picture}(15,9)(2,1)
\includegraphics{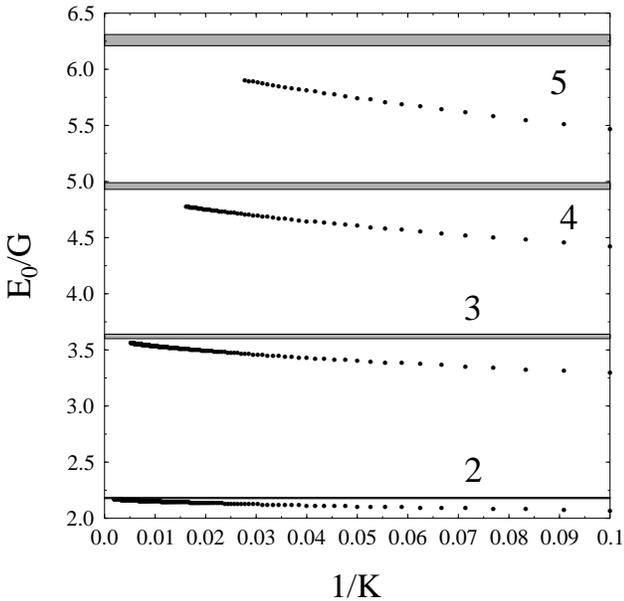}
\end{picture}
\caption{Ground state energy of winding modes for several
transverse sizes $n_\perp$ of the lattice as a function
of the inverse DLCQ parameter $K$. The shaded bands correspond
to the $K\rightarrow \infty$ extrapolated results for the
ground state energies. The width of the bands reflects
systematic uncertainties in the extrapolation.}
\label{fig:string1}
\end{figure}

In the calculation of the glueball masses, we used DLCQ \cite{pa:dlcq}
with anti-periodic boundary conditions, because the convergence
in the DLCQ parameter $K$ is faster than with periodic boundary
conditions. 
Finally, the masses of "wrap-around glueballs" were calculated for
a fixed size of the periodic lattice as a function of the DLCQ
parameter $K$ and the results were extrapolated to $K \rightarrow
\infty$ (Fig.\ref{fig:string1}).
The transverse string tension (in lattice) was then extracted by
considering the resulting extrapolated masses of 
these states as a function of the number of lattice spacings. 
Note that, as the nearly equidistant spacing in Fig.\ref{fig:string1}
indicates, already for only a few lattice
spacings, the mass of these states depends almost exactly
linearly on the number of transverse lattice spacings, which
allows one the easily extract the transverse string tension
from this data.
The fitted result is
\begin{equation}
\frac{M(n)}{G} \stackrel{n\rightarrow \infty}
{\longrightarrow}n\sigma_\perp a \approx
n \left(1.30 \pm 0.03\right)\quad
,
\end{equation}
where we introduced the transverse string tension and the
transverse lattice spacing.
If one denotes the longitudinal string tension by
$\sigma_L$, one finds (\ref{eq:lstring})
\begin{equation}
\sigma_L =G^2 \frac{\pi}{2}
\end{equation}
and if one furthermore imposes $\sigma_L \stackrel{!}{=}
\sigma_\perp =\sigma$
one finds for the transverse lattice spacing in physical units
\begin{equation}
a \quad\stackrel{\mbox{def}}{=}  \quad
\frac{\lim_{n\rightarrow \infty} \frac{M(n)}{n} }{\sigma_\perp}
\approx \frac{1}{G}\left( 0.83\pm 0.02 \right)
\end{equation}
Having related the longitudinal and transverse scales, we are now
in a position to explore the $Q\bar{Q}$ potential.
Note that even though the "wrap-around glueballs" are quite useful
in the determination of the 
purely transverse string tension, they cannot be used to determine
the string tension for any direction other than the transverse
direction and they also cannot be used to determine the
$Q\bar{Q}$ potential for finite separations.
The $Q\bar{Q}$ potential will therefore be determined by using the
fixed charges formalism outlined above.
Nevertheless, since --- as we will see below --- the determination
of the string tension from the $Q\bar{Q}$ potential through the
fixed charges formalism is less accurate than the method
using wrap-around glueballs, we will still use the latter method to
fix the scales in the following calculation.

\begin{figure}
\unitlength1.cm
\begin{picture}(15,9)(2,1)
\includegraphics{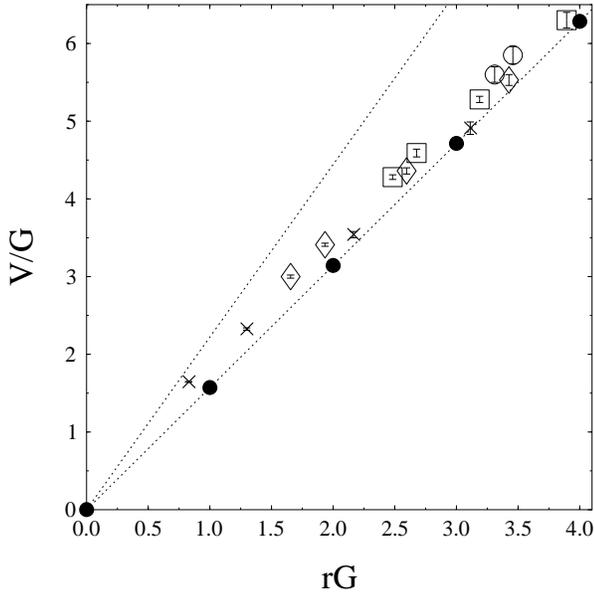}
\end{picture}
\caption{$Q\bar{Q}$-potential versus $r$, where $r^2=x_\perp^2+x_L^2$,
as extracted from the light front Hamiltonian in
the approximation where only one gluon per link is allowed.
The error bars reflect the uncertainties arising from the extrapolation
to the continuum limit in the DLCQ calculation.
The transverse scale was fixed from the masses of "wrap-around glueballs".
The fact that not all points lie on the same smooth curve reflects the
residual anisotropy due to the Fock space truncation and the use of an
oversimplified effective potential.
For comparison, the two dashed lines give the range of values one would obtain
in the large $m$ limit, $V\propto |x_L|+|x_\perp|$.
Dots, crosses, diamonds, squares and circles correspond to $n_\perp=0,1,2,3,4$
respectively.
}
\label{fig:vqq1}
\end{figure}

\begin{figure}
\unitlength1.cm
\begin{picture}(15,10)(-10.7,.5)
\includegraphics{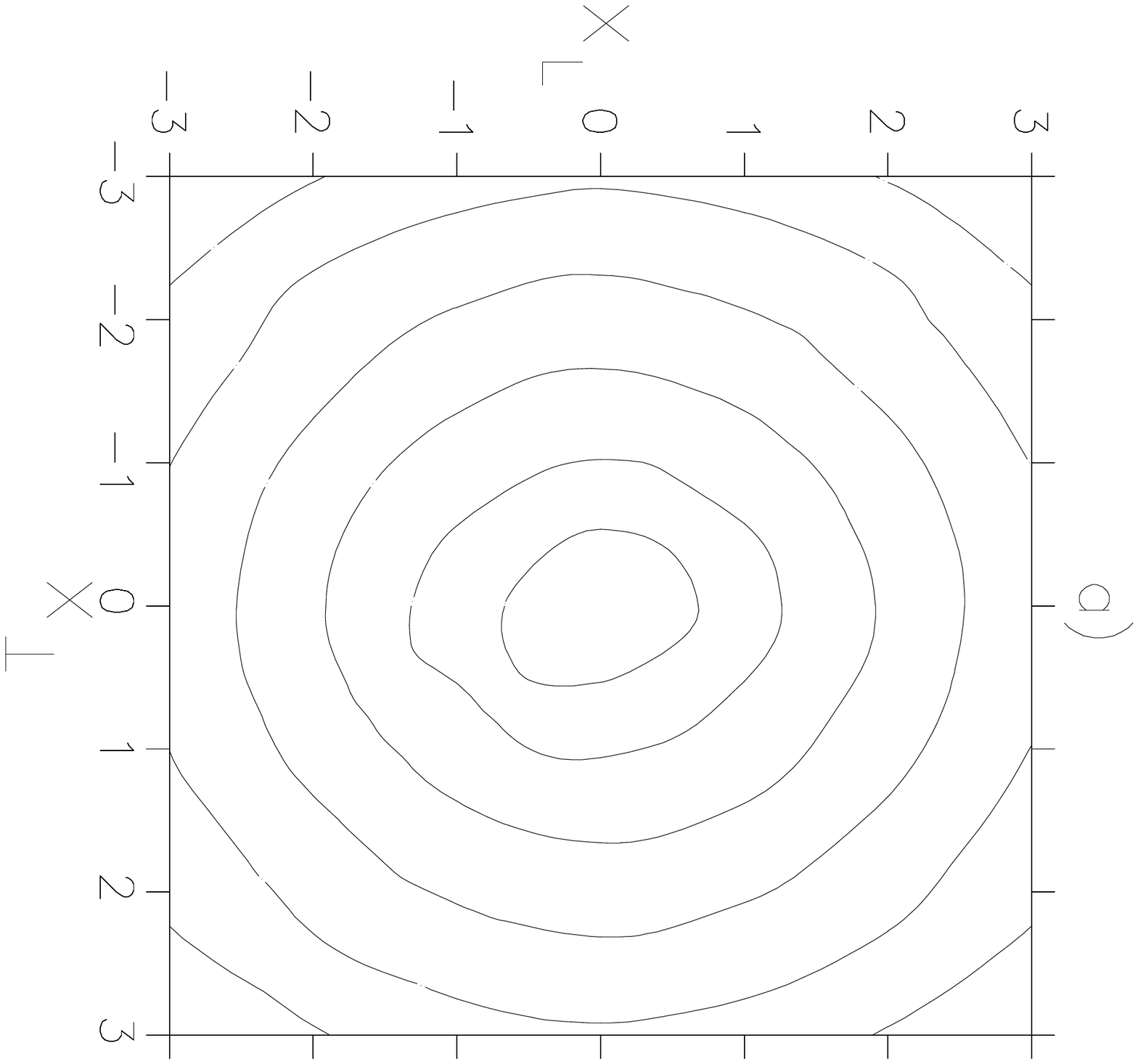}
\end{picture}
\begin{picture}(15,9.5)(-11.,.5)
\includegraphics{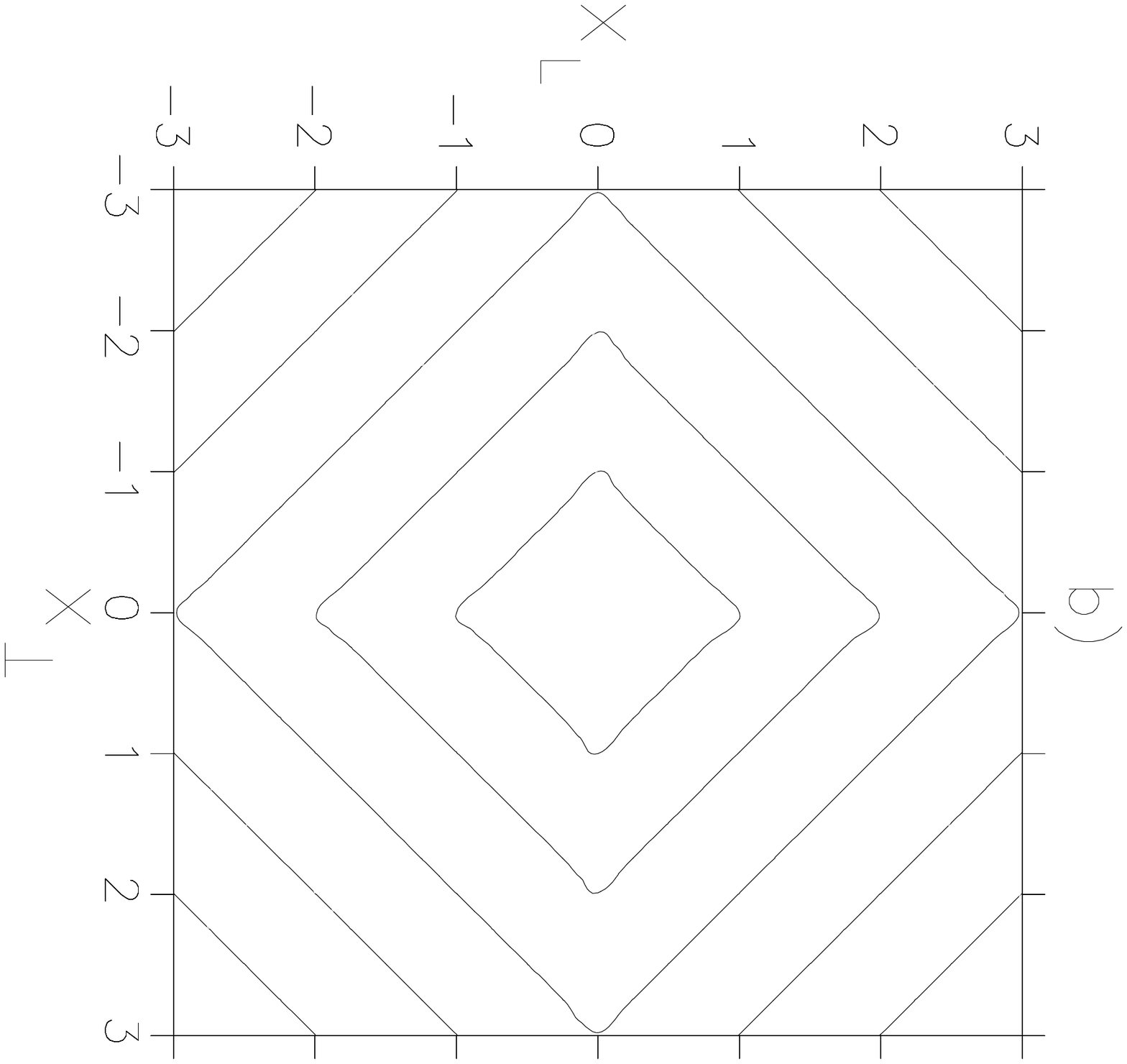}
\end{picture}
\caption{a.) Contour plot for the $Q\bar{Q}$-potential versus $x_\perp$
and $x_L$ in the approximation where we keep only one gluon
per link. For comparison, the contour plot for 
the $Q\bar{Q}$-potential in the large $m$ 
(strong coupling) limit ($V\propto |x_L|+|x_\perp|$)
is shown in b.).}
\label{fig:vqq2}
\end{figure}

In the calculation of the $Q\bar{Q}$ potential we proceeded as
follows: For a given longitudinal separation $x^-$ and a given
velocity $v^+$, the DLCQ Hamiltonian (for a given number of
transverse sites) was constructed with a cutoff on the total
longitudinal momentum of the gluon string\footnote{In contrast
to DLCQ calculations of glueball masses, momentum is not conserved
here since arbitrary momenta can be transferred to and from the
external charges.}
Since the momenta are discrete, $v^+$ provides an infrared
cutoff in the following sense: the peak of the wave function, will be
at momenta of order $k^+ \approx v^+ G$. I.e. only if $v^+G$ is large
in integer units, one will not be affected by the cutoff.
Thus one must perform a careful extrapolation\footnote{This must
be done very carefully because the peak
of the wave function moves as one changes $v^+$.}, were one sends both
the UV cutoff, but also the velocity $v^+$ to infinity. 
The actual numerical work was based on DLCQ with anti-periodic boundary
conditions on the link fields. The ground state energies and wave functions 
for the resulting DLCQ Hamiltonians 
were determined using a Lanzcos algorithm \cite{hiller}.
The resulting $Q\bar{Q}$ potential is shown if Fig.\ref{fig:vqq1}.
Even though not all points in Fig.\ref{fig:vqq1} lie on the same smooth curve,
there is still a significant improvement compared to the large $m$ 
limit
(\ref{eq:vstatic}), where values for $V(x_L,x_\perp)$ would fill the whole area
within the two dashed lines in Fig.\ref{fig:vqq1}.
The residual anisotropy is mostly due to the use of an oversimplified
effective potential [Eq.(\ref{eq:vc2})] but also due to the (ad hoc) suppression
of higher Fock components (see also the results below obtained without
Fock space truncation). Nevertheless, the restoration of rotational
invariance is quite impressive as a contour plot for the same data
as in Fig.\ref{fig:vqq1} shows (Fig.\ref{fig:vqq2}).
The lines of constant potential turn out to be almost circles.
The slight anisotropy of the data in Fig.\ref{fig:vqq1} results in
a slight ellipsoidal distortion in Fig.\ref{fig:vqq2}.
Note that the ellipsoidal shape in Fig.\ref{fig:vqq2} does {\it not}
mean that the string tension in the longitudinal and transverse
directions are different! In fact they are the same by construction,
i.e. by the choice for the lattice unit. What is going on here is
that, while $V(x_L,0) = G^2 \frac{\pi}{2} |x_L|$ in the
approximation chosen in this work, the potential in the transverse 
direction satisfies
$V(0,x_\perp) \stackrel{ x_\perp \rightarrow \infty}{\longrightarrow}
G^2 \frac{\pi}{2} |x_\perp|+c$, which has the same string tension in
longitudinal and transverse direction but which results in a equipotential
lines that look like ellipses.
\section{The Momentum Carried by the Glue}
\label{sec:xglue}
Even though the calculations presented in this paper are still
very crude, the approximate rotational invariance of the 
$Q\bar{Q}$ potential is very encouraging so that we proceed to
investigate other physical observables.
We picked the gluon distribution in these $Q\bar{Q}$ systems since 
one of the main motivations to study light-front QCD in the
first place is the direct access to parton distributions
measured in deep inelastic scattering. In particular, we will
focus on the (light-front) momentum carried by the gluon component.

Before we proceeds, we should caution the reader that
the results must be interpreted with care: Since we work with
finite and large lattice spacing, the "gluons" are not point-like 
"current gluons" and thus the distribution functions should be considered
as distribution functions of some kind of "constituent gluons".
Nevertheless, the results might be helpful in obtaining some
intuitive insight about the parton structure at low $Q^2$.

Obviously, when the external charges are sitting on the same
site, there are no gluons within the approximation
considered in this paper and thus the gluons carry zero
momentum for such a configuration. The other extreme is
when the charges are separated only transversely, in which
case one can derive an exact result:

Consider Eq.(\ref{eq:1site}) for $x^-=0$
\begin{eqnarray}
P^-\psi(k^+) &=& \left( \frac{k^+}{2{v^+}^2}+
\frac{m^2}{2k^+}\right)\psi(k^+)
\label{eq:1site0} 
\\
& &\!\!\!\!\!\!\!\!\!\!\!\!\!\!\!\!\!\!\!\!\!\!\!\!\!\!\!\!\!\!\!+\, G^2\!
\int_0^\infty \!\!\!\!dq^+
\frac{\left(q^++k^+\right)\left[\psi(k^+)-
\psi(q^+)\right]}
{2\sqrt{k^+q^+}\left(q^+-k^+\right)^2}
\nonumber
\\
& &\!\!\!\!\!\!\!\!\!\!\!\!\!\!\!\!\!\!\!\!\!\!\!\!\!\!\!\!\!\!\!+\, G^2\!
\int_0^\infty \!\!\!\!dq^+
\frac{\left(q^++k^+\right)\left[\psi(k^+)-
\psi(q^+)\right]}
{2\sqrt{k^+q^+}\left(q^+-k^+\right)^2}                                           .
\nonumber
\end{eqnarray}
On the one hand (see Appendix A) one knows that
\begin{equation}
P^-=\frac{\langle V\rangle}{v^+},
\end{equation}
where $\langle V\rangle$ is the expectation value for potential (rest-frame) energy in this
configuration. On the other hand,
from the Feynman-Hellman theorem for variations of $P^-$
with respect to $v^+$ in Eq.(\ref{eq:1site0}) one finds
\begin{equation}
\frac{d}{dv^+}P^- = -\frac{1}{{v^+}^3}
\int_0^\infty dk^+ \left| \psi(k^+) \right|^2 k^+.
\end{equation}
Combining these two results, one thus finds the remarkably
simple result (valid for $Q\bar{Q}$ pairs that are separated
only transversely)
\begin{equation}
P^+_{glue} \equiv \int_0^\infty dk^+ \left| \psi(k^+) \right|^2k^+
=\langle V\rangle v^+
\label{eq:sumrule1}
\end{equation}
which states that the momentum carried by the gluons is proportional
to the energy in the gluon field.
Applying the same reasoning to Eq.(\ref{eq:2site}) (again for
$x^-=0$), one obtains similarly
\begin{equation}
P^+_{glue} \equiv \int_0^\infty dk_1^+dk_2^+ \left| \psi(k_1^+,k_2^+) \right|^2\left(k_1^++k_2^+\right)
=\langle V\rangle v^+,
\label{eq:sumrule2}
\end{equation}
which has the same simple interpretation as Eq.(\ref{eq:sumrule1}).
The generalization to more than 2 links is straightforward and
obvious:
\begin{eqnarray}
P^+_{glue} &\equiv& \int_0^\infty dk_1^+...dk_n^+ \left| \psi(k_1^+,...,k_n^+) \right|^2\left(k_1^++...+k_n^+\right)
\nonumber\\
&=&\langle V\rangle v^+.
\label{eq:sumrulen}
\end{eqnarray}
Note that the similar sum rules have been derived elsewhere 
\cite{heft} and they are not limited to the Fock space truncated
version.

\begin{figure}
\unitlength1.cm
\begin{picture}(15,9)(2,1)
\includegraphics{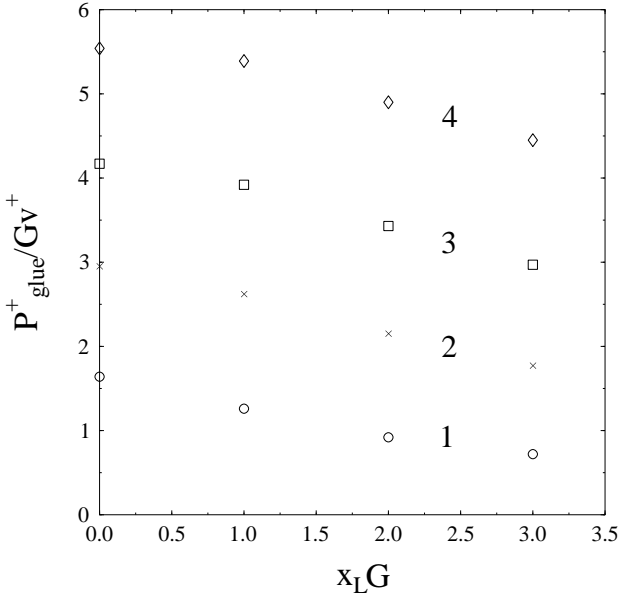}
\end{picture}
\caption{Momentum carried by the gluons for external charges
that are separated by $n_\perp=1,...,4$ lattice spacings
as a function of the longitudinal (rest-frame) separation of the external
charges. The results are obtained in the one-gluon-per-link
approximation. For $n_\perp=0$ the gluons carry no momentum
in this approximation.
}
\label{fig:xglue}
\end{figure}

In the general case, where both $x_L$ and $x_\perp$ are 
nonzero, we were not able to find a generalization of
Eq.(\ref{eq:sumrulen}). However, we can easily use the
numerically obtained states from the calculation of the
ground state energy and just "measure" the momentum carried
by the glue. The result is shown in Fig.\ref{fig:xglue}.
First of all, one notices that the momentum carried by the
gluons increases as the transverse separation increases.
This result is intuitively obvious and happens for
$x_L=0$ in a way that is consistent with Eq.(\ref{eq:sumrulen}).
A much less obvious result is that, for fixed $n_\perp$, the
momentum carried by the gluons {\it decreases} for increasing
longitudinal separation. This result is at first
counter-intuitive because the energy in the gluon field
increases when one increases $x_L$ for fixed $n_\perp$.
However, this apparent paradox gets resolved when one 
recalls that only transverse components of the electric
field contribute to the momentum of the state in the
infinite momentum frame\cite{photon}. 
Assuming (for simplicity) that
the $Q\bar{Q}$ pair is connected by a straight gluon
string, it is obvious that the transverse component
decreases as the string gets rotated into the longitudinal
direction. This also explains why the momentum carried by
the gluons depends not only on the separation of the
$Q\bar{Q}$ pair but also on its orientation. 
As an idealized case, it is instructive to assume that
the $Q\bar{Q}$ pair is connected by a straight string, i.e.
the electric field lines are parallel to the
$Q\bar{Q}$-axis (which has an angle with the longitudinal
direction that we will denote by $\Theta$.
If $|\vec{E}|$ is the field strength in the string then
$|\vec{E}_\perp|=|\vec{E}|\sin \Theta $. In Ref.\cite{photon}
it has been shown that the momentum carried by the gauge
bosons is proportional to the volume integral of 
$\vec{E}_\perp^2$, i.e. one would expect
\begin{equation}
P^+_{glue}/v^+ \propto r\sin^2\Theta =x_\perp \sin \Theta
=\frac{x_\perp^2}{\sqrt{x_\perp^2+x_L^2}},
\end{equation}
which qualitatively explains the decrease of $P^+_{glue}$
for fixed $x_\perp$ with increasing $x_L$.

The qualitative conclusion that one can take from these results
is that for heavy $Q\bar{Q}$ systems, such as $J/\Psi$ or Upsilon mesons,
it is quite natural that the gluons carry a fraction
\begin{equation}
x_{glue}\equiv \frac{P^+_{glue}}{P^+_{total}} 
=c\frac{\langle V\rangle}{M_{Q\bar{Q}}}
\label{eq:xqq}
\end{equation}
of a hadron's momentum, 
where $c$ is a "geometry factor", which is of the order 1.

\section{Including Higher Fock Components}
Even though the results in Section \ref{sec:num} are already amazingly close
to being rotationally invariant, they have a serious flaw: 
they were obtained after an {\it ad hoc} truncation of the
Fock space to the minimal component ($\leq 1$ quanta per link).
While this provided us with a very intuitive and numerically
acceptable "quantum mechanical" approach to the gluon
distribution, it gives rise to results that are still very
close to the strong coupling limit: in fact, even though
the fields may fluctuate longitudinally (which distinguishes
the above valence approximation from the strict strong coupling
limit) QED and QCD still give identical static potentials
at this level. One might thus be sceptical about the results obtained
--- especially about the linearly rising $Q\bar{Q}$ potential.
In this section, we will provide evidence that the results 
obtained for the $Q\bar{Q}$ potential do {\it not} depend very 
much on the valence approximation. 

The main difference between the results in this section and the
results in the previous section is that we allowed for one
additional link-antilink pair in the string state.
Computer storage and the algorithm that we were using
(Lanzcos, where we stored all nonzero matrix elements and their
addresses  in order to keep the code fast)
did not allow us to go beyond one additional
pair while at the same time being able to extrapolate to the
continuum limit for two or more links.
However, our calculations with one additional pair showed that
this component of the wave function is already very strongly
suppressed compared to the valence component so that we do not
expect significant changes of the result by going to even
higher Fock components. Also, we should point out that with
just one additional pair there is a difference between
QED and QCD.

The effective potential was taken to
be the same as the one we used for the valence calculation,
namely just a quadratic term and we did the calculations with
the "mass" at the $1^{st}$ order critical point.
We should emphasize that adding higher order terms to the
effective potential only affects the calculation through higher
Fock components. Therefore (unless the coefficients of the
higher order terms are taken to be very large to invalidate
the perturbative argument) we expect the $Q\bar{Q}$ potential
to be rather insensitive to such higher order terms.
This is in sharp contrast to glueball calculations where even
the ground state
spectrum depends strongly on the higher order terms \cite{bvds}.
The rest of the calculational procedure strongly resembled the
calculation in the previous section. Because of limitations of our
algorithm, we had to restrict ourselves to 3 or less lattice
sites.

\begin{figure}
\unitlength1.cm
\begin{picture}(15,10)(2,.5)
\includegraphics{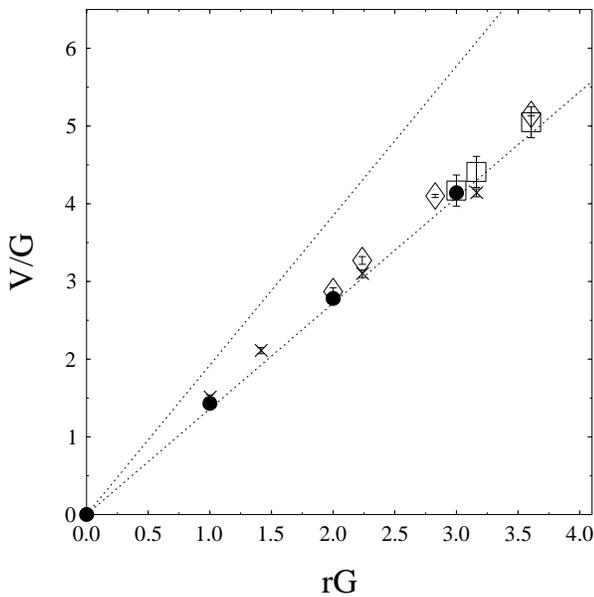}
\end{picture}
\caption{Contour plot for the $Q\bar{Q}$-potential versus $x_\perp$
and $x_L$ in the approximation where we allow for up to one additional
link-antilink pair.}
\label{fig:wqqq}
\end{figure}

\begin{figure}
\unitlength1.cm
\begin{picture}(15,9)(2,1)
\includegraphics{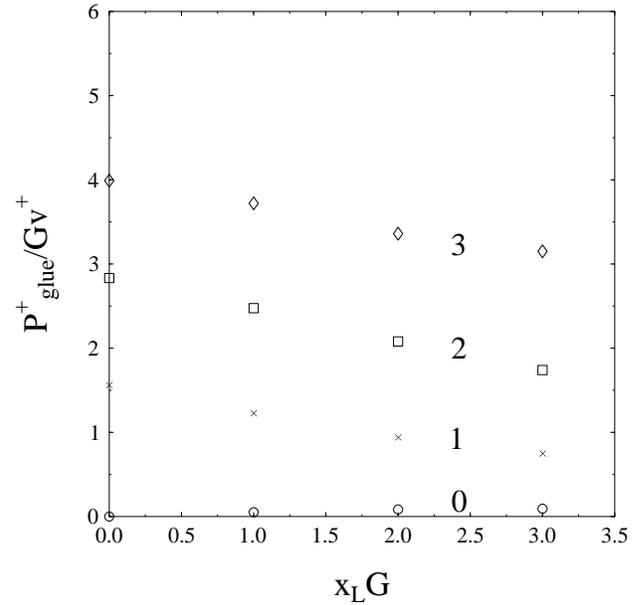}
\end{picture}
\caption{Momentum carried by the gluons for external charges
that are separated by $n_\perp=0,...,3$ lattice spacings
as a function of the longitudinal (rest-frame) separation of the external
charges. The results are obtained in the one-gluon-per-link
approximation plus one additional pair. 
}
\label{fig:wxglue}
\end{figure}

The most important result was that higher Fock components in the
ground state of the string had a norm of typically only a few
per cent, with the resulting energy shifts slightly larger but
still of the same order of magnitude. In fact, this justifies 
the truncation to only one pair. The results for the $Q\bar{Q}$
potential and the momentum carried by the glue are shown in Figs. 
(\ref{fig:wqqq}) and (\ref{fig:wxglue}) respectively.
The first thing that one notes when one compares Fig. \ref{fig:vqq1}
and Fig. \ref{fig:wqqq} is an increase in the size of the error bars.
This is because we had to limit ourselves to smaller values of the
longitudinal momentum and the systematic uncertainties from the
extrapolation increased. Another obvious observation is
that by including an additional pair the result
gets even closer to being rotationally invariant.
We find that very encouraging for future calculations.
Apart from that, the only difference between the calculations with
and without the extra pair is a renormalization of the string tension
in lattice units ($a_\perp = 1.14 \sigma^{-1/2}$ compared to
$a_\perp (valence) = 1.04 \sigma^{-1/2}$).

The momentum carried by the glue no longer has to vanish for
a $Q\bar{Q}$ pair that is oriented longitudinally
(Fig. \ref{fig:wxglue}). Nevertheless,
the numerical result is very small, which reflects the small admixture
from higher Fock components. In fact, similar to the potential,
there is only very little change for the momentum carried by the
glue when one compares calculations with (Fig. \ref{fig:wxglue})
and without (Fig. \ref{fig:xglue}) link-antilink pairs. 
Overall, we found only a small change in the results for the ground state 
of the string as we allowed for higher Fock components.
We find this a very encouraging result since it may allow one to
construct models for quarkonia that have a similar truncation in the
Fock space.

\section{Summary and Outlook}
\label{sec:sum}
We have presented analytical and numerical calculations for the
rest-frame $Q\bar{Q}$ potential in QCD on a transverse lattice
using light-front quantization. The rest-frame $Q\bar{Q}$ potential
was obtained by considering the invariant mass of a heavy $Q\bar{Q}$
pair that moves with constant velocity and fixed separation.

First we showed analytically, that one obtains a linear potential
(both in the longitudinal as well as in the transverse direction)
in the limit of large transverse lattice spacings. This limit is
similar to the strong coupling limit in Hamiltonian or Euclidean
lattice QCD. The confinement mechanism for QCD in LF gauge 
in this limit depends on the orientation of the external charges:
In the longitudinal direction, confinement arises from the
instantaneous interaction which appears when the unphysical
component of the gluon field is eliminated by solving the
LF gauge analog of the Poisson equation. This component of the
gluon field is also unphysical in the sense that it carries
no momentum in the infinite momentum frame. Thus, for a
purely longitudinally separated $Q\bar{Q}$ pair the gluon field
carries no momentum at all in the limit of large transverse
lattice spacings --- even though there is potential energy in
the gluon field. For transversely separated charges, a
completely different confinement mechanism is at work for large
transverse lattice spacings: gauge invariance demands that the
$Q\bar{Q}$ pair is connected by a string of gauge-link 
field.\footnote{Even if one is not dogmatic about gauge invariance,
infrared divergences, that occur otherwise, practically enforce
this condition.} The "strong coupling limit" in Hamiltonian QCD
corresponds on a transverse lattice to the limit, where the
Lagrangian for the gluon link contains a large mass and thus
the ground state energy of a gluon configuration is obtained by
counting links. Together with the instantaneous interaction
between quarks and adjacent link fields as well as among adjacent
link fields, one thus obtains a square shaped linear potential
--- a result that is familiar from the strong coupling limit.
Note that the link fields carry momentum in the infinite
momentum frame, i.e. for transversely separated $Q\bar{Q}$ pairs,
the gluons do carry a sizeable fraction of the total momentum.

Even though such an extreme asymmetry --- no momentum carried by
the gluons for purely longitudinal separations --- 
is the result of the strong coupling limit, it is actually quite
natural and physical that there is such an asymmetry depending
on the orientation. This effect occurs already in QED 
\cite{photon} and arises from the fact that the component of the
(color-) electric which is transverse to the boost direction
transforms differently from the component parallel to it.
In particular, the parallel component does not
contribute to the Pointing vector in the infinite momentum frame.
For this reason, and because of the difference in the field distribution
between QED and QCD (dipole versus string)
we expect a stronger orientational dependence
of the gluon momentum in QCD than in QED.

In the next step we performed numerical calculations of the
$Q\bar{Q}$ potential. First we showed results where we truncated the
Fock space to not more than one quantum of the link field for each
transverse link. Otherwise, the link fields were treated fully
dynamical and were allowed to move freely in the longitudinal
direction. There was only a first
order transition as a function of the link field mass, i.e.
while the transverse lattice spacing in physical units decreased
with the mass term, it did not approach zero at the critical
point. Nevertheless, the numerically obtained
$Q\bar{Q}$ potential at the critical point was almost rotationally
invariant. 
The reason for the first order critical point was most likely 
our oversimplified choice of the effective link field potential
$V_{eff}(U)$. 

Even though the numerical precision was only limited, we then showed that
the qualitative results change only little when we allowed for
excited Fock space components. This is very important, since only
upon inclusion of higher Fock components is there a difference between
QED and QCD. The unimportance of higher Fock components in the ground state
of the string is presumably partly responsible for the approximate rotational
invariance of our results. The reason is that in general one would
not expect rotational invariance on a coarse lattice unless one includes
the full effective potential for the link fields --- from which we kept only
the quadratic term. Those terms that we omitted, contribute only to
matrix elements that include higher Fock components. Therefore, if
higher Fock components are not important for the $Q\bar{Q}$ potential then the
higher order terms in $V_{eff}(U)$ cannot be important either.\footnote{Note that
this argument is not fully complete since we have not verified that the
higher Fock components are unimportant for the ground state of the string
after the higher order terms are included in $V_{eff}(U)$.}
For tests of rotational invariance that are sensitive to the higher order
terms in $V_{eff}(U)$ one must therefore study other observables, such
as glueball spectra\cite{bvds}.

The numerical results, both with and without truncation of the Fock space,
confirmed the strong dependence of the momentum carried by the glue on
the orientation of the external pair. This might have observable consequences
for heavy quarkonia with nonzero orbital angular momentum.

There are many extensions of this work that one could think of, such as
studying excited states of the string or using Monte Carlo algorithms for
obtaining the eigenstates and eigenvalues of the LF Hamiltonian.
Excited states would be interesting both for theoretical but also for
phenomenological reasons. On the theoretical side, since higher Fock components 
play a stronger role in the excited states of the string, they would allow probing
the relevance of the higher order terms in $V_{eff}(U)$ for rotational 
invariance and could thus be used to help fix those constants.
From the phenomenological point of view, the excited states of the string
are interesting because of their connection to hybrid states.
Employing Monte Carlo algorithms\cite{lfepmc} when studying the transverse
lattice might be a useful option since that would allow one to include
many more Fock components than otherwise.
Repeating the calculations in $3+1$ dimensions is an obvious extension
of this work \cite{progress}.  One possibility in this direction, which we
are currently exploring, is to repeat the minimal Fock space truncation
and to fit the plaquette interaction term such that approximate
rotational invariance in all 3 spatial directions is achieved.
Then one can use this gluonic interaction to investigate the physics
of the ground and excited states of heavy quarkonia.

\acknowledgements
M.B. acknowledges useful discussions with
F. Antonuccio, S. Dalley, H.J. Pirner and B.vande Sande.
This work was supported by the D.O.E. under contract DE-FG03-96ER40965
and in part by TJNAF.

\section{Appendix: How to measure the rest-frame $Q\bar{Q}$ 
potential in a light-front calculation}

In this Appendix\footnote{See also Ref.\cite{mb:conf}.}, a scheme
is developed which allows one to extract the $Q\overline{Q}$
potential, i.e. the quantity which corresponds to the potential
between two infinitely heavy quarks in a rest frame, from a
light-front calculation.\footnote{Even though the body of this
paper in on $2+1$ dimensional QCD, we keep the discussion
in this section independent of the number of transverse space-time
dimensions.}
There are several reasons to study this observable in the
light-front (LF)
framework
\begin{itemize}
\item The LF formalism lacks manifest rotationally invariance.
Therefore, if one starts with a {\it wrong} LF Hamiltonian
for QCD, the result $V(\vec{R})$ depends on the orientation
of ${\vec R}$ with respect to the 3-axis. \footnote{We use the
notation $A^\pm = A_\mp = (A^0\pm A^3)/\sqrt{2}$,
${\vec A}_\perp = (A^1,A^2)$.}
A measurement of $V(\vec{R})$ thus provides a direct probe of
rotationally invariance in a physical observable.
\item This sensitivity of $V(\vec{R})$ to rotationally symmetry can then be exploited
in the renormalization procedure to help determine non-covariant
counter terms.
\item  And, most importantly, $V(\vec{R})$ in QCD is very well known over
a large range of distances from the spectroscopy of heavy
$Q\overline{Q}$ mesons as well as from non-perturbative Euclidean
lattice calculations (at least in the absence of dynamical quarks
--- but it is easy to make the same approximation in a LF framework).
\end{itemize}
\unitlength1.1cm
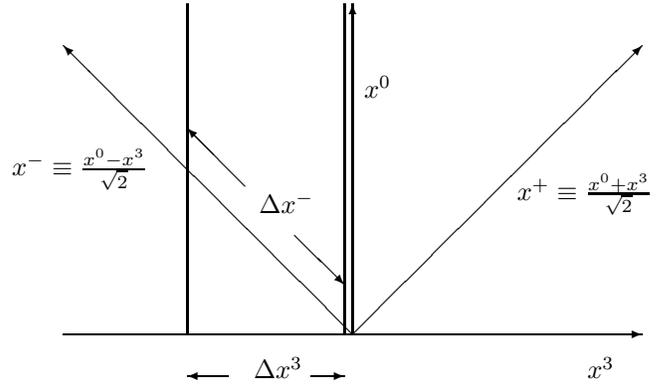
\begin{figure}
\begin{picture}(10.,5.4)(3.3,-1.)
\put(4.,.5){\vector(1,0){7.}}
\put(7.5,.5){\vector(0,1){4.}}
\put(7.5,.5){\vector(1,1){3.5}}
\put(7.5,.5){\vector(-1,1){3.5}}
\put(7.8,3.5){\makebox(0,0){$x^0$}}
\put(10.5,.1){\makebox(0,0){$x^3$}}
\put(10.3,2.2){\makebox(0,0){$x^+\equiv\frac{x^0+x^3}{\sqrt{2}}$}}
\put(4.2,2.5){\makebox(0,0){$x^-\equiv\frac{x^0-x^3}{\sqrt{2}}$}}
\thicklines \put(5.5,.5){\line(0,1){4.}}
\thicklines \put(7.4,.5){\line(0,1){4.}}
\thinlines
\put(6.,0){\vector(-1,0){.5}}
\put(7.,0){\vector(1,0){.4}}
\put(6.6,.1){\makebox(0,0){$\Delta x^3$}}
\put(6.2,2.3){\vector(-1,1){.7}}
\put(6.8,1.7){\vector(1,-1){.6}}
\put(6.7,2.1){\makebox(0,0){$\Delta x^-$}}
\end{picture}
\caption{World lines for two charges with longitudinal
separation $\Delta x^3$ in the rest frame.}
\label{world:fig}
\end{figure}
Before we embark on deriving the effective LF Hamiltonian
for two infinitely heavy sources, it is instructive to
understand physically what it means to have two fixed
sources ``at rest''\footnote{Here we mean ``at rest'' for a
conventional observer, i.e. in an equal time frame.}
from the LF point-of-view.
As should be clear from Fig.\ref{world:fig}, \underline{fixed charges}
in a \underline{conventional frame}
correspond to \underline{charges that move with constant velocity}
on the \underline{LF} [$v^+=v^-=1/\sqrt{2}$ in the example in
Fig.\ref{world:fig}].
Furthermore, if the longitudinal separation is $\Delta x^3$ in
the rest frame, the charges have fixed separation
$\Delta x^-=\sqrt{2}\Delta x^3$ in the
longitudinal LF direction. In the more general case, where the charges are moving
with constant four velocity $v^\mu$, where ${\vec v}_\perp=0$
in the rest frame, one obtains $\Delta x^- = \Delta x^3/v^+$.

The transverse
separation is the same on the LF as it is in a rest frame
description.
Therefore, in order to understand the LF-physics of charges
that are fixed in a conventional frame
with separation ${\vec R}=(R^1,R^2,R^3)$,
we must first understand how to describe a ``dumbbell,''
with ends separated by $(\Delta x^1,\Delta x^2,\Delta x^-)
=(R^1,R^2,R^3/v^+)$,
that moves
with constant velocity $v^+$.

\subsection{One Heavy Quark on the LF}

A pair of sources, moving both with the same constant velocity
$v^\mu$ can also be interpreted (and treated)
as {\it one extended source} moving with constant velocity $v^\mu$
(for simplicity, we will keep ${\vec v}_\perp ={\vec 0}$).
This is reminiscent of heavy quarks and thus, as a warmup
exercise, it is very instructive to consider one (point-like) heavy
quark on the LF first (see also Refs.\cite{mb:conf,quatschek}).

For simplicity, we will first take the heavy quark limit
for the canonical Hamiltonian, which can be written in the form
\begin{equation}
P^-_B = \frac{M_b^2+{\vec k}_{b\perp}^2}{2p_b^+}
+P^-_{HL}+P^-_{LL},
\label{eq:pminus2}
\end{equation}
where $B$ represents the hadron, $b$ is the heavy quark,
$P^-_{HL}$ contains the interactions between heavy ($b$) and
light degrees of freedom and $P^-_{LL}$ contains all
terms involving light degrees of freedom only.
The heavy quark limit is obtained by making an expansion in
inverse powers of the $b$-quark mass. For this purpose we write
\begin{equation}
p^+_b=P^+_B - p^+_L=M_Bv^+-p_L^+,
\label{eq:P+}
\end{equation}
where $p_L^+$ is defined to be the sum of the longitudinal
(LF-) momenta of all light degrees of freedom. For the (total)
LF-energy we write on the l.h.s. of Eq.(\ref{eq:pminus2})
\begin{equation}
P^-_B=M_Bv^-=\frac{M_B}{2v^+}= \frac{M_b+\delta E}{2v^+},
\label{eq:P-}
\end{equation}
where $\delta E \equiv M_B-M_b$ is the ``binding energy'' of
the hadron.
After inserting Eqs.(\ref{eq:P+}) and (\ref{eq:P-}) into Eq.(\ref{eq:pminus2})
and expanding one obtains
\begin{eqnarray}
\frac{M_b+\delta E}{2v^+}
&=&
\frac{M_b^2 + {\vec k}_{b\perp}^2}{2(M_Bv^+-p_L^+)} +P^-_{HL} +
P^-_{LL}
\label{eq:mexpand}
\\
&=&\frac{M_b^2 + {\vec k}_{b\perp}^2}{2\left(M_bv^++\delta E v^+-p_L^+\right)}
+P^-_{HL} + P^-_{LL}
\nonumber\\
&=&
\frac{M_b}{2v^+} - \frac{\delta E}{2v^+}+ \frac{p_L^+}{2v^{+2}}
+{\cal O}\left(\frac{1}{M_b}\right)+P^-_{HL} + P^-_{LL}.
\nonumber
\end{eqnarray}
Note that we have assumed that the transverse momentum of the
heavy quark is small compared to its mass, which is justified
in a frame where the transverse velocity of the heavy
hadron vanishes.
The term proportional to $M_b$ cancels between the l.h.s.
and the r.h.s. of Eq.(\ref{eq:mexpand}) and we are left with
\begin{equation}
\frac{\delta E}{v^+}= \frac{p_L^+}{2v^{+2}}
+{\cal O}(1/M_b)+P^-_{HL} + P^-_{LL}.
\label{eq:deltae}
\end{equation}
The {\it brown muck} Hamiltonian $P^-_{LL}$ is the same as
for light-light systems and will not be discussed here.
The interaction term between the heavy quark and the
brown muck ($P^-_{HL}$) is more tedious but straightforward.
For example, heavy quark pair creation terms (via instantaneous
gluons) are proportional to $1/(p^+_{b_1}+p^+_{b_2})^2
\propto 1/M_b^2$ and can thus be neglected. Similarly, pair
creation of heavy quarks from virtual gluons is also suppressed
by at least one power of $M_b$. This also justifies our omission
of states containing more than one heavy quark from the
start.
Other terms that vanish
in $P^-_{HL}$ include interactions that involve
{\it instantaneous exchanges} of heavy quarks, which are
typically proportional to the inverse $p^+$ of the
exchanged quark and thus of the order ${\cal O}(M_b^{-1})$.
Up to this point, all interaction terms that we have
considered vanish in the heavy quark limit. The more interesting
ones are of course those terms which survive.
The simplest ones are the instantaneous gluon exchange
interactions with light quarks or gluons, which are,
respectively,
$V_{Qq}\propto (p_q^+-p_q^{'+})^{-2}$ and
$V_{Qg}\propto (p^+_g+p'^+_g){(p_g^+-p'^+_g)}^{-2}$ and
remain unchanged in the limit $M_b\rightarrow \infty$.
Terms which involve instantaneous gluon exchange and are
off-diagonal in the brown muck Fock space behave in the same
way.

The quark gluon vertex simplifies considerably. For finite
quark mass one has for the matrix element for the emission of
a gluon with momentum $k$, polarization $i$ and color $a$ between quarks
of momentum $p_1$ and $p_2$
\begin{equation}
P^-_{QQg} = -igT^a \left\{
2\frac{k^i}{k^+} -
\frac{ {\vec \sigma}_\perp {\vec p}_{2\perp}-iM_b}{p_2^+}
\sigma^i  -\sigma^i \frac{ {\vec \sigma}_\perp {\vec p}_{1\perp}+iM_b}{p_1^+}
\right\}
,
\end{equation}
where spinors as well as creation/destruction operators have been
omitted for simplicity. In the heavy quark limit [note
$1/p_1^+ -1/p_2^+ = {\cal O}(M_b^{-2})$] the spin dependent terms
drop out and one finds in the heavy quark limit
\begin{equation}
P^-_{QQg} = -2igT^a \frac{k^i}{k^+} .
\end{equation}
The spin of the heavy quark thus decouples completely, giving rise to
the well known $SU(2N_f)$ symmetry in heavy quark systems.

\subsection{Two Heavy Sources}

As we discussed above, two heavy sources at fixed
separation can formally be treated as one extended heavy source
\footnote{Just think of a dumbbell.}.
Therefore, the LF Hamiltonian for two sources is the same as for one
source with two minor modifications:
\begin{itemize}
\item All vertices involving a heavy source get modified
according to the rule (
${\hat O}$ stands for any operator
acting on the brown muck)
\begin{eqnarray}
&&\!\!\!\!\!\!\!\!\!\! \left\{
{\chi^\dagger}' T^a \chi \,\times {\hat O}
\right\} + h.c.\longrightarrow \quad\quad \quad\quad \quad\quad
\quad\quad
\\
&&\!\!\!\!\!\!\!\!\!\!\left\{
\left[ {\chi_Q^\dagger}' T^a \chi_Q F_R(q)
-\chi_{\bar{Q}}^\dagger T^a {\chi_{\bar{Q}}}' F_R(-q)
\right] \times {\hat O} \right\}  + h.c.
,
\nonumber
\end{eqnarray}
where $\chi$ stands for the operator acting on the
color degrees of freedom of the heavy quark/antiquark.
The ``form factor''
\begin{equation}
F_R(q)=\exp\left[\frac{i}{2}\left(\frac{q^+ R^3}{v^+}-{\vec
R}_\perp{\vec q}_\perp\right)\right]
\label{eq:form}
\end{equation}
arises from acting with the (kinematic !) displacement
operator on the position of the heavy quark/antiquark 
(shifting it from $x^-=0$, ${\vec
x}_\perp = {\vec 0}_\perp$ to $x^-=\pm R^3/2v^+$, ${\vec x}_\perp =
\pm {\vec R}_\perp/2$)
and $q$ is the {\it net} momentum transferred to the brown muck.
This rule holds irrespective of the number of gluons involved in
this process.
Note that ${\hat O}( brown\, muck)$ is the same for one or two heavy sources.
\item There is a static potential between the two heavy quarks.
In the continuum, the canonical Hamiltonian yields
$P^-_{HH} = g^2\,{\chi_Q^\dagger}' T^a \chi_Q\,
\chi_{\bar{Q}}^\dagger T^a
{\chi_{\bar{Q}}}'\, \delta^{(2)}({\vec R}_\perp)|R^3|v^+$.
In general, there will be a more complicated dependence on ${\vec R}$
which has to be determined by demanding self-consistency.
For example, singularities arising from exchange of gluons
with low $q^+$ between the two sources should cancel (nonperturbatively)
with the IR behavior of the instantaneous potential in $P_{HH}^-$ 
\cite{brazil}.
\end{itemize}
Using Eq.(\ref{eq:deltae}), $V({\vec R})$ can thus be extracted as follows:
\begin{itemize}
\item[1.] For a given ${\vec R}$ and $v^+$, write down the effective
LF-Hamiltonian for the heavy pair interacting with the brown muck
(including the form factors Eq.(\ref{eq:form}) and including all
the counter terms and counter term functions which would also appear
in a ``heavy-light'' system).
\item[2.] The lowest eigenvalue $\delta E^{(1)}$ from Eq.(\ref{eq:deltae}),
i.e. the QCD-ground state in the presence of the two heavy sources,
is then equal to $V({\vec R})$.
\end{itemize}
The heavy quark potential thus calculated is equivalent to
the potential which a lattice theorist would extract from
an asymmetric rectangular Wilson loop. Since there is plenty
of ``quenched'' lattice data around, it would make sense to
omit light quarks completely in a first approach and to
focus on the pure glue part of the brown muck. However, the
formalism described above is so general that one could also
use it in a LF calculation that includes (dynamical) light quarks.

\end{document}